\documentclass[prl,aps,epsfig,showpacs]{revtex4}
\usepackage{bm}
\usepackage{amsfonts}
\usepackage[dvips]{graphicx}
\usepackage{mathrsfs}
\usepackage[intlimits]{amsmath}
\usepackage[colorlinks, citecolor=red]{hyperref}

\begin{document}

\title{Experimental demonstration of a quantum router}
\author{X. X. Yuan$^{1}$, J.-J. Ma$^{1}$, P.-Y. Hou$^{1}$, X.-Y. Chang$^{1}$%
, C. Zu$^{1}$, L.-M. Duan$^{2,1}$\footnote{%
Email: lmduan@umich.edu.}}
\affiliation{$^{1}$Center for Quantum Information, IIIS, Tsinghua University, Beijing 100084,
PR China}
\affiliation{$^{2}$Department of Physics, University of Michigan, Ann Arbor, Michigan 48109, USA}
\date{\today}

\begin{abstract}
The router is a key element for a network. We describe a scheme to
realize genuine quantum routing of single-photon pulses based on cascading
of conditional quantum gates in a Mach-Zehnder interferometer and report a
proof-of-principle experiment for its demonstration using linear optics
quantum gates. The polarization of the control photon routes in a coherent
way the path of the signal photon while preserving the qubit state of the
signal photon represented by its polarization. We demonstrate quantum nature
of this router by showing entanglement generated between the initially
unentangled control and signal photons, and confirm that the qubit state of
the signal photon is well preserved by the router through quantum process
tomography.
\end{abstract}

\maketitle

\section{Introduction}

Quantum network has many potential applications \cite{1,2}. A key element to
build a network is the router, which uses a control bit to determine the
path of the signal bit. In a quantum router, both the control and the signal
bits are represented by quantum bits in general in superposition states, and
the control bits should have the ability to control the paths of the signal
bits in a quantum coherent way \cite{r1,r2}. Such quantum coherent routing
of signal bits offer new remarkable opportunities compared with its
classical counterpart \cite{r1,r2}. For instance, the quantum routing
operation provides the key element to realize the quantum random access
memory \cite{r4}, an essential component for large scale quantum computation
based on the von Neumann architecture and quantum machine learning that
deals with large sets of data \cite{r5,r6}.

In a quantum network, the signal is usually carried by single-photon pulses,
which are ideal realization of the flying qubits for long-distance
communication \cite{1,2}. Several experiments have demonstrated routing of
single-photon pulses when the control bit takes only classical states \cite%
{5,6}. For instance, an optical switch can efficiently route single-photon
pulses based on micro-electromechanical or optical control \cite{5,6}. In a
cavity QED (quantum electrodynamic) system, single trapped atoms or
superconducting circuits are able to route the path of single photons \cite%
{7,8,9a,9b,9c}. In a genuine quantum router, the control bit may take
quantum superposition states to route the paths of the signal photon in a
coherent way. At the same time, the signal photon, apart from its path, need
to have another degree of freedom to carry its qubit state (quantum data),
which should be preserved by the quantum routing operation \textbf{\cite{r2}}%
. Such a device acts like a quantum transistor, performing entangling gates
on the paths of the single-photon pulses while preserving their qubit states
\textbf{\cite{1,2}}. The ability to coherently route the path of the signal
bits is critical for realization of the quantum random access memory \cite%
{r4}. So far no experiments have demonstrated full quantum nature of a
router. The cavity QED system in principle can be used to realize a genuine
quantum router \cite{r4}. However, this requires precise coherent control of
both the matter and the photonic qubits, which is experimentally challenging
despite the recent remarkable advance \cite{9a,9b,9c}.

In this paper, we report a proof-of-principle demonstration of genuine
quantum routing of single-photon pulses, where the control signal,
represented by the polarization of a single photon, can take arbitrary
superposition states. We demonstrate the key features of a quantum router
\textbf{\cite{r1}}: (1) the control and the signal photons take independent
input quantum states, and the polarization of the control photon routes in a
coherent way the path of the signal photon, generating polarization-path
entanglement between the initially unentangled control and signal photons;
(2) The qubit state of the signal photon represented by its polarization is
well preserved by the router, so the routing operation does not destroy
quantum data carried by the signal pulse. We first describe a general scheme
that can realize deterministic quantum routing based on cascading of two
quantum CNOT\ gates in a Mach-Zehnder interferometer. Deterministic quantum
CNOT\ gates between single-photon pulses require large nonlinearity at the
single-photon level, which remains experimentally daunting. Instead, in our
specific demonstration, we use post-selected quantum gates based on linear
optical elements \cite{10} and realize quantum routing in a probabilistic
fashion based on post-selection by the coincidence measurements \cite%
{10,11,12}. We unambiguously confirm the characteristic quantum features of
the router through quantum state and process tomography. Cascading of
post-selected linear optics gates has been reported before for realization
of the $3$-bit Toffoli gate \cite{11a}, but the scheme is different for
implementation of the quantum routing operation. The implementation based on
the post-selected gates can not be scaled up to many qubits. However,
similar to linear optics quantum computation \cite{3,4}, we can in principle
make the gate and routing scheme more scalable by combining linear optics elements with
feed-forward from the high-efficiency single-photon detection.

\section{Results}

\subsection{An implementation scheme for quantum routing}

The idea of a quantum router is illustrated in Fig. 1(a). The signal photon
need to have two degrees of freedom: polarization and path. Its polarization
is used to carry the quantum data, represented by a qubit state $\left\vert
\Phi \right\rangle _{s}=d_{0}\left\vert H\right\rangle _{s}+d_{1}\left\vert
V\right\rangle _{s}$ with arbitrary coefficients $d_{0},d_{1}$, where $%
\left\vert H\right\rangle _{s}$ and $\left\vert V\right\rangle _{s}$ denote
two orthogonal linear polarizations. The incoming path of the signal photon
is denoted by $\left\vert U\right\rangle _{s}$. After the quantum router,
the outgoing path of the signal photon is determined by an address qubit,
which is represented by the polarization state of a control photon. For a
classical router (optical switch), the outgoing path of the signal photon is
either $\left\vert U\right\rangle _{s}$ or $\left\vert D\right\rangle _{s}$,
determined by the polarization of the control photon which takes either $%
\left\vert H\right\rangle _{c}$ or $\left\vert V\right\rangle _{c}$. For a
quantum router, the control qubit is in a quantum superposition state $%
\left\vert \Psi \right\rangle _{c}=c_{0}\left\vert H\right\rangle
_{c}+c_{1}\left\vert V\right\rangle _{c}$ with arbitrary coefficients $%
c_{0},c_{1}$, and coherence between the two classical routing possibilities
should be maintained. So the routing operation generates a polarization-path
entangled state $\left\vert \Psi \right\rangle _{cs}=c_{0}\left\vert
H\right\rangle _{c}\left\vert U\right\rangle _{s}+c_{1}\left\vert
V\right\rangle _{c}\left\vert D\right\rangle _{s}$\ between the initially
unentangled control and signal photons. Such path entanglement from quantum
routing is a key requirement for realization of the quantum random access
memory \cite{r4,r5,r6}. Similar to a classical router, which does not
destroy the data carried by signal photon, a quantum router should preserve
the polarization state $\left\vert \Phi \right\rangle _{s}$ that encodes the
quantum data. So, after an ideal quantum router, the final state of the
system takes the form%
\begin{equation}
\left\vert \Psi _{f}\right\rangle =\left( c_{0}\left\vert H\right\rangle
_{c}\left\vert U\right\rangle _{s}+c_{1}\left\vert V\right\rangle
_{c}\left\vert D\right\rangle _{s}\right) \otimes \left( d_{0}\left\vert
H\right\rangle _{s}+d_{1}\left\vert V\right\rangle _{s}\right) .
\end{equation}%
The quantum routing operation transforms the polarization and the path
degrees of freedom of the signal photon in different ways, performing
effectively a quantum CNOT gate on the path of the the signal photon while
preserving its polarization state. This poses a challenge for the
experimental realization as conditional quantum gates between the photons
are typically on the polarization degrees of freedom \cite{10,11,12,11a}.
Note that the definition of the quantum routing operation here is somewhat
different from the one in Ref. \cite{r1}, where after the routing the
control photon and the signal photon are not in an entangled state. The
polarization-path entanglement between the control and the signal photons is
a key element in our approach to quantum routing as this provides the
critical resource for its application in realization of quantum random
access memory \cite{r4} and in achievement of exponential speedup in large
data processing \cite{r5,r6}.

We first describe a general scheme to realize quantum routing based on
cascading of two quantum CNOT\ gates in a Mach-Zehnder interferometer as
shown in Fig. 1(b). A signal photon, initially in the polarization state $%
\left\vert \Psi _{s}\right\rangle $ with arbitrary superposition
coefficients $d_{0},d_{1}$, is incident from one side of the polarization
beam splitter (PBS) of the Mach-Zehnder interferometer. The PBS correlates
the photon's polarization and path degrees of freedom and transforms its
state to $d_{0}\left\vert HU\right\rangle _{s}+d_{1}\left\vert
VD\right\rangle _{s}$. The control photon, initially in the state $%
\left\vert \Psi _{c}\right\rangle $, meets the signal photon successively
through the paths $D$ and $U$. When the two photons meet each other, we
perform a quantum CNOT\ gate which flips the polarization of the signal
photon if and only if the control photon is in $H$-polarization. The Pauli
operation $X$ in the circuit exchanges polarization $H$ and $V$ for the
photon in the corresponding path. After the second PBS\ of the Mach-Zehnder
interferometer, the output state is given by $\left\vert \Psi
_{f}\right\rangle $ (see Supplementary Material for detailed derivation). So
the optical circuit in Fig. 1(b) achieves exactly the quantum routing
operation. The scheme described here is simpler and more general than the
one proposed in Ref. \cite{r1}: first, we don't need the quantum
non-demolition measurement required in \cite{r1}. Furthermore, this implementation
scheme, by itself, is deterministic, not limited to linear optics gates, and
applies to any experimental systems where one can realize quantum CNOT\
gates on single-photon pulses.

\begin{figure}[tbp]
\includegraphics[width=8.5cm,height=7cm]{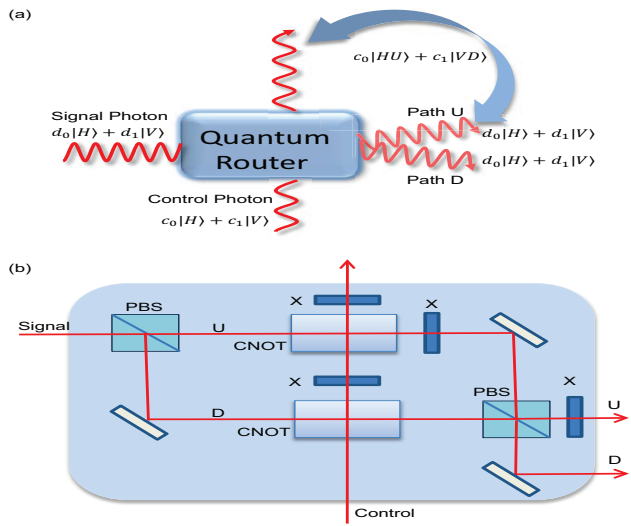}
\caption[Fig. 1 ]{\textbf{Principle and Scheme for a quantum router.}
\textbf{a}, Illustration of the principle of a quantum router. A control
qubit, represented by the polarization state of a single photon, routes the
output paths of the signal photon in a coherent way, generating
polarization-path entanglement between the initially unentangled control and
signal photons. The quantum data, carried by the polarization state of the
signal photon, is preserved by the routing operation. \textbf{b}, A scheme
to implement the quantum routing operation through an optical circuit with
quantum CNOT\ gates, X gates, and an Mach-Zehnder interferometer.}
\end{figure}

\subsection{Experimental setup}

The proposed scheme in Fig. 1(b) for realization of the quantum router is
deterministic if one can realize deterministic CNOT\ gates. As a
proof-of-principle experiment, here we demonstrate a probabilistic version
of this scheme using the post-selected linear optics quantum CNOT gates \cite%
{10}, with the experimental setup shown in Fig. 2. This experimental setup
has some similarity with the one in Ref. \cite{12aa} recently exploited for
implementation of quantum state fusion. However, there is an important
difference: the setup in Ref. \cite{12aa} does not generate polarization-path
entanglement between the control and the signal photons, which is a key
feature of our quantum router scheme. As first demonstrated in Ref. \cite{10}, the
optical circuit shown in Fig. 2(b) realizes a post-selected CNOT\ gate on
the two input modes, conditional on the case that one photon exits from each
of the output modes, which occurs with a probability of $1/9$ \cite{10,12a}.
In our experimental setup shown in Fig. 2(c), the CNOT\ gate is combined
with the Pauli gate $X_{c}$ on the control photon, so the last half wave
plate (HWP) at $45^{o}$ can be removed. The control photon after the first
CNOT\ gate needs to go through the second CNOT\ gate in the upper arm of the
Mach-Zehnder interferometer. The combined success probability of these two
successive gates is $1/27$, corresponding to the case of one photon in the
control mode and the other photon in one of the interferometer arms of the
signal mode (see the Supplementary Material for details on post-selection
measurements). After the quantum router, we confirm quantum coherence
between the two output paths of the signal photon through another
Mach-Zehnder interferometer as shown in Fig. 2(c).

\begin{figure}[tbp]
\includegraphics[width=14cm,height=8cm]{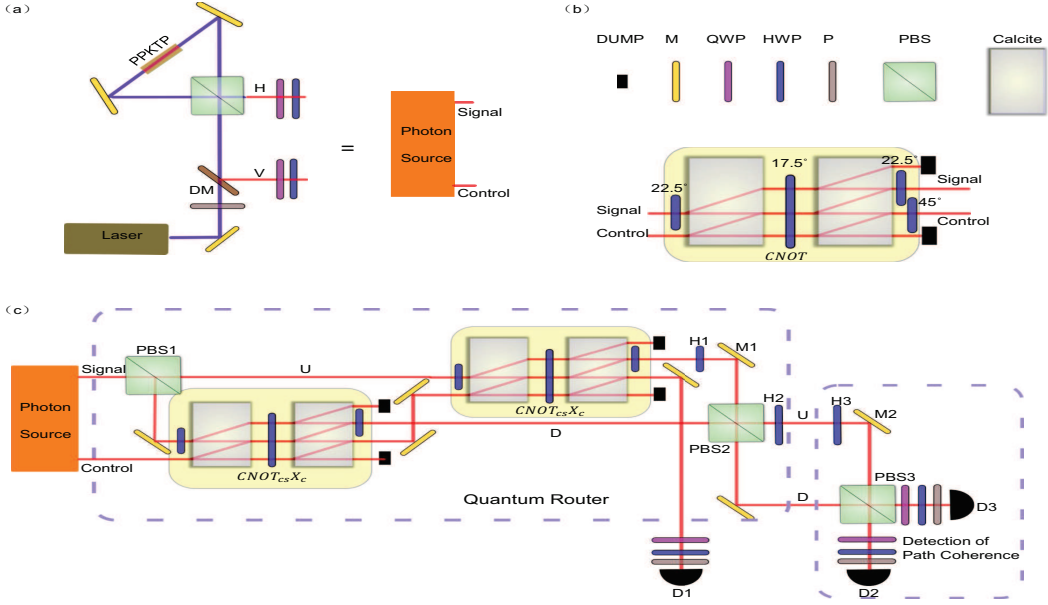}
\caption[Fig. 2 ]{\textbf{Experimental setup for realization of a quantum
router.} \textbf{a}, Experimental setup for the photon source to generate
control and signal photons with independently controlled arbitrary
polarization states. \textbf{b}, Experimental setup for a post-selected
quantum CNOT\ gate between the photon and the signal photons, where the
numbers show the corresponding angles of the half wave plates (HWPs).
\textbf{c}, Experimental setup for the quantum router. The quantum CNOT\
gate $CNOT_{cs}$, with the signal photon as the target qubit, is combined
with the X-gate $X_{c}$ on the control photon, and realized by the optical
elements shown in the yellow-shaded boxes. The two big Mach-Zehnder
interferometers, one by PBS1 and PBS2, and the other by PBS2 and PBS3, are
both actively phase locked to maintain phase stability, where the locking
laser beams and optical/electronic devices are not shown in the figure for
clarity of the picture (see Methods for details). The HWPs H1 and H2 are
both set at an angle of $45^{o}$, performing X gates on the photon's
polarization in the corresponding path. The optical elements in the last
dash-line box is only required for detection of coherence between the output
paths ($U$ or $D$) of signal photon. The HWP H3 is also set at $45^{o}$. The
rotation of HWPs, quarter wave plates (QWPs), and polarizers (P) before each
of the single-photon detectors (D1, D2, and D3), combined together, can
choose an arbitrary polarization basis for photon detection, which is
required for quantum state tomography. Tomographic measurement of quantum states is performed
by registering the two-photon coincidence events between the detectors D1 and either D2 or D3.}
\end{figure}

\subsection{Experimental results}

To demonstrate quantum routing operation, first we rotate the polarization
state $\cos \theta \left\vert H\right\rangle _{c}+\sin \theta \left\vert
V\right\rangle _{c}$ of the control photon by continuously varying $\theta $%
, and check the output path of the signal photon prepared in $H$%
-polarization. The recorded coincidence counts for the signal photon in the $%
U$ or $D$ path are shown in Fig. 3(a), which follow oscillation curves $\cos
^{2}\theta $ and $\sin ^{2}\theta $, respectively. The oscillation of photon
counts in the $U$ path has a high visibility of $97.0\%$, while the
corresponding visibility for the $D$ path is only $84.6\%$. The difference
in visibility comes from different roles played by the CNOT gates for this
scenario: as one can check from the optical circuit in Fig. 2(c), for the
signal photon to go to the $D$ path, the CNOT\ gate plays an active role to
flip its polarization through two-photon Hong-Ou-Mandel interference, which
typically has a larger imperfection \cite{10}. Such a polarization flip is
not required for the signal photon to go to the $U$ path. The difference of these two
visibilities of about $12\%$ is therefore a characterization of the imperfection of the
linear optics CNOT gate. The infidelity of the CNOT gate is mainly induced by the imperfect mode matching between different optical
paths for the Hong-Ou-Mandel interference. In Fig. 3(b), we
show the path information of signal photon now prepared in $V$-polarization.
We observe similar oscillation curves except that the visibilities for the $%
U $ and the $D$ paths are exchanged (of $80.0\%$ and $97.3\%$ respectively)
as the role of CNOT\ gate is reversed in this case.

\begin{figure}[tbp]
\includegraphics[width=8.5cm,height=10cm]{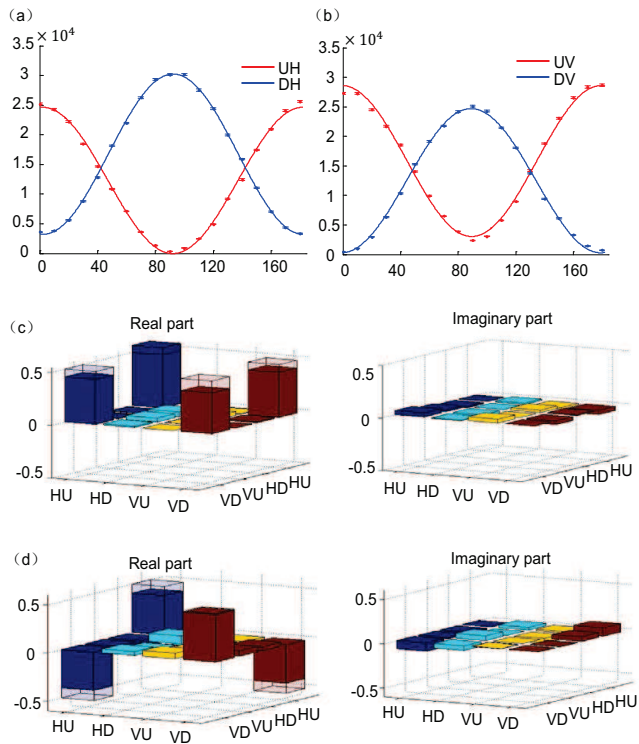}
\caption[Fig. 3 ]{ \textbf{\ Experimental verification of a quantum router.}
\textbf{a}, The detected coincidence photon counts (accumulated over $13$
seconds) in different paths are shown as functions of the polarization angle
$\protect\theta$ of control photon. The angle $\protect\theta$ varies from $%
0^{o}$ to $180^{o}$, with the corresponding polarization state $\cos(\protect%
\theta)\left\vert V \right\rangle_c + \sin(\protect\theta)\left\vert H
\right\rangle_c $. The symbols UH and DH mean that the signal
photon is in U or D path with H-polarization, and the corresponding data represent
the photon coincidence counts between the detectors D1 and either D2 or D3.
The error bars denote standard
derivation and their calculation is specified in detail in Methods. \textbf{b%
}, Same as Fig. a, but with the signal photon now in $V$-polarization.
\textbf{c}, Real and imaginary parts of the reconstructed density matrix
elements for the polarization-path state of control and signal photons. The
input polarization state of control photon is $\left( \left\vert
H\right\rangle _{c}+\left\vert V\right\rangle _{c}\right) /\protect\sqrt{2}$
and of signal photon is $\left\vert H\right\rangle _{s}$. The hollow caps
denote the matrix elements for the ideal output state $\left( \left\vert
H\right\rangle _{c}\left\vert U\right\rangle _{s}+\left\vert V\right\rangle
_{c}\left\vert D\right\rangle _{s}\right) /\protect\sqrt{2}$ after a perfect
router. \textbf{d}, Same as Fig. c, but now the input polarization state of
control photon is $\left( \left\vert H\right\rangle _{c}-\left\vert
V\right\rangle _{c}\right) /\protect\sqrt{2}$ and of signal photon is $%
\left\vert V\right\rangle _{s}$.}
\end{figure}

To confirm quantum nature of this router, we demonstrate coherence between
the routing paths and polarization-path entanglement generated between the
control and the signal photons initially in product states. We set the input
polarization state $\left\vert \Psi _{c}\right\rangle $ of the control
photon to $\left( \left\vert H\right\rangle _{c}+\left\vert V\right\rangle
_{c}\right) /\sqrt{2}$. In the idea case, the output for the control and the
signal photons should be in the polarization-path maximally entangled state $%
\left( \left\vert H\right\rangle _{c}\left\vert U\right\rangle
_{s}+\left\vert V\right\rangle _{c}\left\vert D\right\rangle _{s}\right) /%
\sqrt{2}$. We verify this entanglement by reconstructing the density matrix
for the output polarization-path state through measurements based on quantum
state tomography \cite{13}. For two-qubit states, the quantum state
tomography is done with $16$ independent measurements in complementary bases
and the density matrix is reconstructed using the maximum likelihood method
\cite{13}. The real and imaginary parts of all the elements of density
matrix are shown in Fig. 3(c) with the signal photon carrying $H$%
-polarization state. From this measurement, we find that the entanglement
fidelity $F$ of the output state is $F=(88.5\pm 0.5)\%$, well above the
boundary of criterion $F>0.5$ for demonstration of entanglement \cite{18}.
The entanglement is also confirmed for other superposition states of $%
\left\vert \Psi _{c}\right\rangle $ and found to be almost independent of
the polarization state of signal photon. As an example, in Fig. 3(d) we show
the reconstructed density matrix elements for the output polarization-path
state with $\left\vert \Psi _{c}\right\rangle =\left( \left\vert
H\right\rangle _{c}-\left\vert V\right\rangle _{c}\right) /\sqrt{2}$ and the
signal photon in $V$-polarization. The corresponding entanglement fidelity
in this case is $F=(83.0\pm 0.5)\%$. The small decrease in entanglement
fidelity (about $5\%$) results from the slightly larger imperfection for one of the CNOT\
gates in the Mach-Zehnder interferometer, and this fidelity difference is consistent with the difference in the two visibilities
($84.6\%$ versus $80.0\%$) that we observed in Fig. 3(a) and 3(b).

The quantum router should preserve quantum data carried by the polarization
state of signal photon. Although the polarization of signal photon plays an
important role in the Mach-Zehnder interferometer and in quantum CNOT gates,
the whole quantum routing operation combining all the elements together
should not change this polarization state. So, in the subspace of quantum
data represented by the polarization qubit of signal photon, the router just
performs an identity gate. To confirm this, we reconstruct the routing
transformation in this subspace through quantum process tomography (see
Methods) \cite{19}. The reconstructed process matrix elements are shown in
Fig. 4 with the control photon initially in $H$-polarization. From the
result, we conclude that the process fidelity $F_{P}=\left( 91.9\pm
0.3\right) \%$ and the average gate fidelity $\overline{F}=\left( 94.6\pm
0.2\right) \%$. For the $V$-polarization component of control photon, the
measured process matrix in the quantum data subspace is very similar and
thus not shown in the figure. We find the corresponding average gate
fidelity for this latter case is $\overline{F}=\left( 92.7\pm 0.2\right) \%$.

\begin{figure}[tbp]
\includegraphics[width=8.5cm,height=4cm]{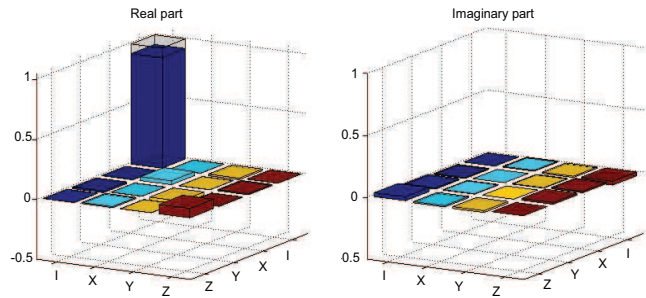}
\caption[Fig. 4 ]{\textbf{Quantum process tomography of the data qubit.}
Real and imaginary part of the reconstructed process matrix elements for the
data qubit carried by the polarization state of signal photon. The quantum
router preserves the data qubit, so in the ideal case the operation is
represented by an identity operator with its elements shown by the hollow
caps. The figure shows the case with the control photon in $H$-polarization.
For $V$-polarization, the figure looks pretty much the same.}
\end{figure}

\section{Discussion}

Experimental demonstration of a quantum router opens up prospects for its
application in quantum networks and quantum data processing. It provides the
key element for realization of the quantum random access memory \cite{r4}.
As the paths of signal photons get entangled with the control qubits in a
quantum router, it allows us to make a delayed choice of the routing
destination, either to one location or superposition of several locations,
similar to the quantum delayed-choice experiments \cite{r3}. Such a
delayed-choice routing, apart from its fundamental interest, may have
applications in network cryptography. The control qubits perform effectively
entangling gates on the paths of quantum signals. In analogy to quantum
parallelism in superfast quantum algorithms, such entangling gates may be
exploited for achieving constructive interference between the paths of
signals, leading to quantum parallel distribution of signals in a network.

\textbf{Acknowledgements} This work was supported by the National Basic
Research Program of China 2011CBA00302. LMD acknowledges in addition support
from the IARPA MUSIQC program, the AFOSR and the ARO MURI program.

\textbf{Author Contributions} L.M.D. designed the experiment and supervised
the project. X.X.Y., J.J.M., P.Y.H., X.Y.C., C.Z. carried out the
experiment. L.M.D. and X.X.Y. wrote the manuscript.


\section{Methods}

\subsection{\protect\bigskip Derivation of quantum routing operation through
the circuit in Fig. 1(b)}

The input of the control and the signal photons is given by a product state $%
\left\vert \Psi _{in}\right\rangle =\left( c_{0}\left\vert H\right\rangle
_{c}+c_{1}\left\vert V\right\rangle _{c}\right) \otimes \left(
d_{0}\left\vert H\right\rangle _{s}+d_{1}\left\vert V\right\rangle
_{s}\right) \left\vert U\right\rangle _{s}$ with arbitrary coefficients $%
c_{0},c_{1},d_{0},d_{1}$. After the first PBS\ of the Mach-Zehnder
interferometer, the state transforms to
\begin{equation*}
\left\vert \Psi _{1}\right\rangle =c_{0}d_{0}\left\vert H\right\rangle
_{c}\left\vert H\right\rangle _{s}\left\vert U\right\rangle
_{s}+c_{0}d_{1}\left\vert H\right\rangle _{c}\left\vert V\right\rangle
_{s}\left\vert D\right\rangle _{s}+c_{1}d_{0}\left\vert V\right\rangle
_{c}\left\vert H\right\rangle _{s}\left\vert U\right\rangle
_{s}+c_{1}d_{1}\left\vert V\right\rangle _{c}\left\vert V\right\rangle
_{s}\left\vert D\right\rangle _{s}.
\end{equation*}%
Then, after the CNOT\ gate at the lower interferometer arm and an $X$ gate
on the control photon, the state becomes%
\begin{equation*}
\left\vert \Psi _{2}\right\rangle =c_{0}d_{0}\left\vert V\right\rangle
_{c}\left\vert H\right\rangle _{s}\left\vert U\right\rangle
_{s}+c_{0}d_{1}\left\vert V\right\rangle _{c}\left\vert H\right\rangle
_{s}\left\vert D\right\rangle _{s}+c_{1}d_{0}\left\vert H\right\rangle
_{c}\left\vert H\right\rangle _{s}\left\vert U\right\rangle
_{s}+c_{1}d_{1}\left\vert H\right\rangle _{c}\left\vert V\right\rangle
_{s}\left\vert D\right\rangle _{s}.
\end{equation*}%
The next operation is the CNOT\ gate at the upper interferometer arm
followed by an $X$ gate on the control photon and an $X$ gate on the signal
photon in the $U$ path. After this operation the state evolves to

\begin{equation*}
\left\vert \Psi _{3}\right\rangle =c_{0}d_{0}\left\vert H\right\rangle
_{c}\left\vert V\right\rangle _{s}\left\vert U\right\rangle
_{s}+c_{0}d_{1}\left\vert H\right\rangle _{c}\left\vert H\right\rangle
_{s}\left\vert D\right\rangle _{s}+c_{1}d_{0}\left\vert V\right\rangle
_{c}\left\vert H\right\rangle _{s}\left\vert U\right\rangle
_{s}+c_{1}d_{1}\left\vert V\right\rangle _{c}\left\vert V\right\rangle
_{s}\left\vert D\right\rangle _{s}.
\end{equation*}%
The last operation is the second PBS of the interferometer acting on the
signal photon and an $X$ gate on the signal photon in the $U$\ path. This
gives the final state

\begin{eqnarray*}
\left\vert \Psi _{f}\right\rangle &=&c_{0}d_{0}\left\vert H\right\rangle
_{c}\left\vert H\right\rangle _{s}\left\vert U\right\rangle
_{s}+c_{0}d_{1}\left\vert H\right\rangle _{c}\left\vert V\right\rangle
_{s}\left\vert U\right\rangle _{s}+c_{1}d_{0}\left\vert V\right\rangle
_{c}\left\vert H\right\rangle _{s}\left\vert D\right\rangle
_{s}+c_{1}d_{1}\left\vert V\right\rangle _{c}\left\vert V\right\rangle
_{s}\left\vert D\right\rangle _{s} \\
&=&\left( c_{0}\left\vert H\right\rangle _{c}\left\vert U\right\rangle
_{s}+c_{1}\left\vert V\right\rangle _{c}\left\vert D\right\rangle
_{s}\right) \otimes \left( d_{0}\left\vert H\right\rangle
_{s}+d_{1}\left\vert V\right\rangle _{s}\right) ,
\end{eqnarray*}%
which is exactly the output state one expects from an ideal quantum router.

\subsection{Experimental details}

In our experiment, the photon source is given by a pair of single photons
generated through spontaneous parametric down conversion in a nonlinear
periodically-poled potassium titanyl phosphate (PPKTP) crystal of $15$ mm
length (see Fig. 2a). The crystal is pumped by a continuous-wave (cw) diode
laser at $404$ nm wavelength, generating signal and control photons at $808$
nm wavelength. The sagnac loop interferometer shown in Fig. 2a can generate
polarization entangled photon pairs if the pump beam is set at $\left\vert
H\right\rangle +\left\vert V\right\rangle $ polarization. For our
experiment, however, we set the pump beam at $\left\vert H\right\rangle $
polarization by the polarizer before the dichromatic mirror (DM) so that the
down-converted photons have $H$ and $V$ polarization, respectively, and go
out along different paths after the PBS. The half wave plate (HWP) and the
quarter wave plate (QWP) in the two output paths then prepare the control
and the signal photons respectively to independent polarization states with
arbitrary coefficients $c_{0},c_{1},d_{0},d_{1}$.

The quantum CNOT gate in our experiment, shown in Fig. 2b, is constructed
using the same method as in ref. \cite{10}. It is made of two parallel placed
Calcites with a HWP between them set at an angle of $17.5^{o}$. Calcites
work as a PBS, and the two parallel Calcites make a Mach-Zehnder
interferometer of two loops with intrinsic phase stability. With two-photon
Hong-Ou-Mandel interference in this interferometer and the HWPs set at
appropriate angles given in Fig. 2b, one can check that the device performs
a quantum CNOT\ gate on the incoming control and signal photons, provided
that one post-selects the case that one photon exits from each of the two
output paths \cite{10,11}. The post-selection is done through photon
coincidence measurement. There could be small phase difference in the
optical paths of the Mach-Zehnder interferometer formed by the Calcites,
which is compensated afterwards by adjusting the tilting angle of a
birefringent BBO\ crystal (not shown in the figure). The post-selected
quantum CNOT\ gate is subject to leakage error, where both the control and
the signal photons go to one of the output paths and the final state leaks
outside of the qubit Hilbert space \cite{10,11}. This leakage error leads to
background noise in the coincidence measurement on the outputs of the router
shown in Fig. 2c. The contribution of this background noise is measured in
our experiment by registering the coincidence while blocking either the
control photon path or the signal photon paths after the first CNOT gate (at
the lower arm of the interferometer). The background counts can then be
subtracted from the total counts, which removes this noise. When all the
coefficients $c_{0},c_{1},d_{0},d_{1}$ are nonzero, interference terms may
arise between the background noise and the signal terms. To remove
contribution of these interference terms, we can insert a phase shifter to
the path of the control photon after the first CNOT\ gate to induce a phase
shift of either $\alpha $ ($\alpha $ can be any real number) or $\alpha +\pi
$ per photon. After averaging of the coincidence counts for these two rounds
of experiments with a relative phase shift of $\pi $, the noise interference
terms are removed while the signal terms remain unaffected.

The quantum router setup shown in Fig. 2c has a big Mach-Zehnder
interferometer of long arms that are subject to phase fluctuation, so the
interferometer needs to be actively phase locked to maintain phase
stability. A diode laser beam at $780$ nm wavelength is incident from the
top side of PBS1, goes through the interferometer loop, and is then
separated by a dichromatic mirror after the PBS2 and detected by
photon-detectors. The detected signal, after a PID circuit, provides the
feedback to fine tune the position of mirror M1 in the interferometer loop
through a piezo to maintain stability of the relative phase. After the
router, to detect phase coherence between the two output paths of signal
photon, we combine the paths through another PBS (PBS3). The path coherence
is confirmed by detecting the polarization states of two outputs of the PBS3
in different bases, where the basis selection is achieved through a
polarizer together with a HWP and a QWP. The detected counts by the
single-photon detectors are registered through a coincidence circuit,
recording a signal only when one control photon and one signal photon are
detected. All the error bars in this paper result from the statistical error
associated with the photon detection under the assumption of a Poissonian
distribution for the photon counts. The error bars are propagated from the
registered photon counts to the measured quantities (such as the density
matrix elements and the entanglement fidelity) through exact numerical
simulation. The PBS2 and PBS3 in Fig. 2c make another Mach-Zehnder
interferometer, which requires similar phase locking, achieved through
feedback to the piezo-controlled position of mirror M2 by detection of
another diode laser beam going through this interferometer.

Theoretically, the success probability of the quantum routing operation
shown in Fig. 2 using the post-selected quantum gates is $1/27\approx 0.0370$%
. To experimentally measure this success probability, we first set the angle
of the middle HWP in the CNOT\ optical circuit of Fig. 2(b) to $0^{o}$. At
this angle, both the\ circuit in Fig. 2(b) and the quantum routing circuit
in Fig. 2(c) reduce to a deterministic identity operation on the input
photons. We record the photon coincidence counts in this case as the
comparison point, which is about $6.0\times 10^{4}$ per second. Then we set
the HWPs back to the right angle ($17.5^{o}$) for the quantum routing
operation and measure the coincidence counts of the output photons, which is
$2.2\times 10^{3}$ per second. The ratio between these two coincidence
counts gives the experimentally measured success probability, which is $%
0.0367$, in agreement with the theoretical value. \newline
\qquad For each data point shown in Figs. 3 and 4, we typically collect more
than $10^{5}$ coincidence counts in $13$ seconds. The error bars are
determined by assuming a Poissonian distribution for the photon counts. We
propagate the error bars from the detected photon counts to the quantities
shown in Fig. 3 and 4 through exact numerical simulation by Monte Carlo
sampling according to the Poissonian distribution of the photon counts.

\subsection{Quantum process tomography}

A quantum process is described by a completely positive map $\varepsilon $
which transfers arbitrary initial states $\rho _{i}$ to the corresponding
final states $\rho _{f}\equiv \varepsilon (\rho _{i})$. In quantum process
tomography (QPT), a fixed set of basis operators $\{E_{m}\}$ are chosen so
that the map $\varepsilon (\rho _{i})=\sum_{mn}E_{m}\rho _{i}E_{n}^{\dagger
}\chi _{mn}$ is identified with a process matrix $\chi _{mn}$.\ We
experimentally reconstruct the process matrix $\chi $ through the maximum
likelihood technique \cite{19}. For the single-bit QPT, we choose the basis
operators as $I=I$, $X=\sigma _{x}$, $Y=-i\sigma _{y}$, $Z=\sigma _{z}$ ,
which requires measurement on four different initial states $|H\rangle $, $%
|V\rangle $, $|+\rangle \equiv (|H\rangle +|V\rangle )/\sqrt{2}$, and $%
|M\rangle \equiv (|H\rangle -i|V\rangle )/\sqrt{2}$. To experimentally
measure the process matrix elements shown in Fig. 4, we first set the
polarization of the control photon to $|H\rangle $ and prepare the
polarization of the data photon to one of the four states $\left\{ |H\rangle
,|V\rangle ,|+\rangle ,|M\rangle \right\} $. For each input state, we
measure the output density matrix elements of the data photon by detecting
the photon counts in three complementary bases $\left\{ |H\rangle ,|V\rangle
\right\} $, $\left\{ |+\rangle ,|-\rangle \right\} $, and $\left\{ |P\rangle
,|M\rangle \right\} $, respectively, where $|-\rangle \equiv (|H\rangle
-|V\rangle )/\sqrt{2}$ and $|P\rangle \equiv (|H\rangle +i|V\rangle )/\sqrt{2%
}$. The corresponding output density matrices are determined through the
maximum likelihood method as it is standard for quantum state tomography
\cite{13}. From the measured output density matrices for each of the four
input states $\left\{ |H\rangle ,|V\rangle ,|+\rangle ,|M\rangle \right\} $,
we reconstruct the experimental process matrix $\chi _{e}$ following the
standard maximum likelihood method for quantum process tomography \cite{19},
which gives the data shown in Fig. 4. By the same method, we have also
measured the process matrix for the data photon when the polarization of the
control photon is set to $|V\rangle $ . Finally, we compare the
experimentally reconstructed process matrix $\chi _{e}$ with the ideal one $%
\chi _{id}$ by calculating the process fidelity $F_{P}=Tr(\chi _{e}\chi
_{id})$. The process fidelity $F_{P}$ is connected with the average gate
fidelity $\overline{F}$ through the formula $\overline{F}=(dF_{P}+1)/(d+1)$
\cite{19}, where $\overline{F}$ is the output state fidelity averaged over
all possible input states with equal weight and $d$ is the Hilbert space
dimension ($d=2$ for a single qubit).

\end{document}